\documentclass[submission%
]{dmtcs}

\usepackage[latin1]{inputenc}
\usepackage{subfigure}

\usepackage[round]{natbib}

\newtheorem{proposition}{Proposition}
\newtheorem{lemma}{Lemma}
\newtheorem{definition}{Definition}
\newtheorem{theorem}{Theorem}
\newtheorem{corollary}{Corollary}
\newtheorem{remark}{Remark}
\newtheorem{conjecture}{Conjecture}
\def\P{{\mathbb P}}

\def\N{{\mathbb N}}

\def\mean{\mathbb{E}}
\def\cal{\mathcal}
\def\ind{\mathbf{1}}

\def\ind{\bf{1}}

\RCSdef$Revision: 1.3 $\endRCSdef
\rcsMajMin

\author{Hanène Mohamed\addressmark{1}\thanks{INRIA, France} }

\title{A probabilistic analysis of a leader election algorithm}
\address{\addressmark{1}INRIA Rocquencourt, Domaine de Voluceau, BP 105,
  F-78153 Le Chesnay, France }

\revision{\rcsMaj}

\received{01 April 2006}
\revised{\today}
\accepted{tomorrow}

\begin{document}
\maketitle
\begin{abstract}
A {\em leader election} algorithm is an elimination process that
divides recursively into tow subgroups an initial group of \begin{math}n\end{math}
items, eliminates one subgroup and continues the procedure until a subgroup is of size \begin{math}1\end{math}.
In this paper the biased case is analyzed. We are interested in the {\em cost} of the algorithm, \textit{i.e.} the number of operations needed until the algorithm stops. Using
a 
probabilistic approach,  the asymptotic  behavior of
the algorithm is shown to be related to the behavior of a hitting time of two random sequences on \begin{math}[0,1]\end{math}. 

\end{abstract}

\clearpage
\tableofcontents
\section{Introduction}
\label{sec:in}

A single-hop network is a distributed system of
\begin{math}n\end{math} nodes, also called {\em stations}, sharing a
  common communication channel which can transmit only one message per
  time unit. In the special case of {\em collision} detection,  the channel
  is ternary feedback; each station sending a message to the network
  can simultaneously listen to the channel and  detect: a {\em
  collision} when at least there are two broadcast attempts, a {\em
  silence} when no station sends message, or a {\em success} when
  exactly one station sends its message. A single-hop network with {\em collision} detection is called {\em multiple
  access channel}.

\medskip
\noindent
Consider a {\em multiple access channel} of \begin{math}n\end{math}
  stations which has to elect a {\em
  leader} to control and organize the network. Because of links or stations failures, the {\em leader} may be
temporarily out of service. Such failure can
be detected by a {\em silence}, in which case the system stops normal
operations and initiates the
{\em election} process: the system has to identify a new {\em
  leader} in a reasonable execution time. We are interested in the {\em cost} of the algorithm, i.e. the number of operations needed to find a {\em leader}.

\subsection{Leader election problem}
We assume that the size \begin{math}n\end{math} of the {\em multiple access channel} is unknown. Moreover, each station is assumed to have a unique identifying number ID. To elect a {\em leader}
  among themselves, stations have to use the same algorithm. The case \begin{math}n \in \{0,1\}\end{math} is trivial, \begin{math}n\end{math} is assumed to be greater than \begin{math}2\end{math}. Let us recall the basic one:
\begin{itemize}
\item[---] Deterministic Initialization: At the first time unit, each station send a message with its ID number to the common
				channel. As \begin{math}n
				\geq2\end{math}, all stations detect a {\em collision}. 
\item [---] Randomized Selection Process: Each station \begin{math}S\end{math} generates
  independently a  {\em Bernoulli} random variable \begin{math}
  B_S\end{math} with parameter \begin{math} p \end{math}. Only which
  obtains \begin{math} B_S=1 \end{math} is allowed to send
  its message again during the next time unit. 
\end{itemize}
For a station \begin{math}S\end{math}, there are two cases:
\begin{enumerate}
\item If \begin{math}B_S=1\end{math}, station \begin{math}S\end{math}
    will be called {\em Active}; \begin{math}S\end{math} sends again its message to the channel and can detect
\begin{itemize}
\item[---] a {\em success}; only station \begin{math}S\end{math} is
  trying transmission, then all the other stations receive its ID's message and \begin{math}S\end{math} obtains the status of {\em leader}. The protocol is finished.
\item[---] a {\em collision}; station \begin{math}S\end{math} is not
    the only candidate to be {\em leader}, and so has to generate \begin{math}B_S\end{math} again.
\end{itemize} 
\item Otherwise, station \begin{math}S\end{math} becomes {\em Non Active}; it remains candidate to be {\em leader}, listens to the channel but does not participate to the transmission. So it can detect
\begin{itemize}
\item[---] a {\em success}; only one station \begin{math}S' \not =S\end{math} is trying transmission. The other stations
  (including \begin{math}S\end{math}) detect its ID. So
  \begin{math}S'\end{math} obtains the status of {\em leader}. The protocol is finished.
\item[---] a {\em collision}; although station \begin{math}S\end{math}
  is not participating to the selection process, there are at least  \begin{math}2\end{math} {\em Active} stations. So, station \begin{math}S\end{math} is eliminated.
\item[---] a {\em silence}; all stations are {\em Non Active}, so
  station \begin{math}S\end{math} has to generate
  \begin{math}B_S\end{math} again to send or not its ID's message to the channel.
\end{itemize}
\end{enumerate}
That is, at the end of the protocol, a single station remains {\em Active} and becomes the {\em leader} of the system.

\medskip
\noindent
This {\em splitting} process using a {\em Bernoulli} random variable was also used in the {\em tree protocol} of Capetanakis and Tsybakhov. For a survey, see ~\cite{Mathys:1}. 

\medskip
\noindent
The example below illustrates the election process applied to a group of \begin{math}4\end{math} stations \begin{math}\{A, B, C, D\}\end{math}. In this case, the {\em leader} \begin{math}A\end{math} is elected in \begin{math}4\end{math} times units.

\vspace{0.5cm}

\begin{tabular}{|l|l|l|l|p{2cm}|}
\hline
\verb!time units! & 1 & 2 & 3 & 4\\
\hline
\verb!Active Stations! & A B C D & A B C &  & A\\
\hline
\verb!Non Active Stations! &  & D & A B C & B C \\
\hline
\verb!Eliminated Stations! &  &   &   D   & D   \\
\hline
\verb!Channel feedback! & {\em Collision}   & {\em Collision}   & {\em
  Silence} & {\em Success}  \\ 
\hline
\end{tabular}

\begin{figure}[htbp]
  \begin{center}
    \includegraphics[width=0.5\textwidth]{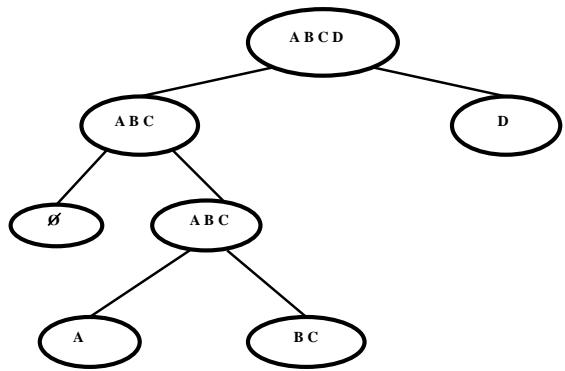}
    \caption{Election of the {\em leader} A; $H_4=
    4$. Incomplete tree structure.}
    \label{fig:logo}
  \end{center}
\end{figure}

\begin{definition}[Algorithm Cost]
 It is the number of rounds needed to find a {\em leader}. Denote by \begin{math}H_n\end{math} the algorithm cost when the size of the network is \begin{math}n\end{math}.
\end{definition}

\medskip
\noindent
Such a randomized elimination algorithm has various applications in
distributed systems like cellular phones and wireless communication
networks. In mobile Ad-hoc networks, failures occur when mobile nodes
move out of transmission range. The unstable topology of the network
makes {\em leader election} problem more complex. For more details, see ~\cite{Malpani:1}. Electing a {\em leader} in a computer network is fundamental
to supervise communication and synchronization. See ~\cite{Fill:1}. It
is also studied in a context of radio networks. For an interesting  survey
on randomized communication in this context, see
~\cite{Chlebus:1}. For more elaborate {\em leader election} algorithms
on radio network with no {\em collision} detection, see ~\cite{Lavault:1}.

\subsection{Splitting process and tree structure}
Formally, the algorithm starts with a group of $n$ items which is divided in
two subgroups. The probability that an item is sent into the left
subgroup is \begin{math}p\end{math}. This subgroup will be divided by the same process. The other items will be ignored. 
If the left subgroup is empty, the algorithm restarts from the previous level.

\medskip
\noindent
This distributed algorithm is a randomized elimination process with a
natural binary tree structure (Fig.\ref{fig:logo}). At the root of the associated tree, is the initial group of items. In the first split, it generates two nodes: the left one will be split by the same process, the right one is a terminal node, also called {\em leaf}, which will not be treated by the algorithm except when the left node is empty. Only in this case, the right node will be split into two one.

\medskip
\noindent
Thus, this tree structure can be represented as an incomplete tree in which only one side is developed. We define the {\em height} of the associated tree as the length of the path from the root to the {\em leader} which is the longest {\em root-to-leaf} path in the tree (see ~\cite{Fill:1} ). Then the algorithm {\em cost} is equivalently the {\em height} of the associated tree. Fig.\ref{fig:logo} illustrates this equality.

\subsection{Previous works}
It is known that the average {\em cost} of the {\em leader election} algorithm is of logarithmic order in \begin{math}n\end{math} with an oscillating behavior. See~\cite{Prodinger:1} for the unbiased case \begin{math}p=1/2\end{math}, ~\cite{Janson:1} for the biased one  \begin{math}p \not=1/2\end{math}.

\medskip
\noindent
Consider the {\em Poisson} model of the {\em leader election} problem, i.e. the {\em election} process applied to a network with random size following a {\em Poisson} process \begin{math}{\cal N}_x\end{math} (see \ref{sec:notations}). Let \begin{math}h\end{math} the {\em Poisson transform} of the sequence of average {\em cost} of the algorithm  \begin{math}\left(\mean(H_n)\right)_{n\geq 0}\end{math}. 
\begin{definition}[{\em Poisson transform}]\label{def}
For \begin{math}x>0 \end{math}, the Poisson transform of the sequence \begin{math}(\mean(H_n))\end{math} is the function  \begin{math}h\end{math} defined by
\[
h(x)=\mean(H_{{\cal N}_x})=\sum_{n=2}^{+\infty}\mean(H_n)\frac{x^n}{n!}~e^{-x}.
\]
\end{definition}
Then, function \begin{math}h\end{math} is solution of a functional equation, called {\em basic functional equation} associated to the algorithm
\begin{equation}\label{functeq}
h(x)=h(px)+h(qx)\, e^{-px}+f(x),~\mbox{ where \begin{math}p+q=1 \end{math}}
\end{equation}
and \begin{math}f \end{math} is a given function. Equation (\ref{functeq}) is the starting point of all studies made on this algorithm.
\subsubsection*{The unbiased case}
When the splitting process follows a {\em Bernoulli} random variable of parameter \begin{math}1/2 \end{math}, the {\em leader election} algorithm is called {\em symmetrical}. Observe that, for the unbiased case, the {\em functional equation} (\ref{eq:functeq})  is solved by direct iteration. In fact, the {\em Poisson transform} \begin{math} h \end{math} verifies
\[
h(x)=h(x/2)\left(1+e^{-x/2}\right)+f(x),
\]
which can be rewritten
\begin{math}
 g(x)=g(x/2)+f(x)/(1-e^{-x})\end{math} where \begin{math}g(x)=h(x)/\left(1-e^{-x}\right)\end{math}.

\medskip
\noindent
The first analysis of the {\em leader election} algorithm was proposed
by ~\cite{Prodinger:1}. He investigated different parameters of interest such as the {\em height}, called {\em depth} in his paper, the {\em size} of the associated tree, \textit{i.e.} the number of nodes\ldots.
Using combinatorial techniques, he established exact expressions and asymptotic formulas for these quantities for the symmetrical case. So, it is shown that for an initial group of size \begin{math} n \end{math}, the algorithm stops on average after about  \begin{math} \log_2 n \end{math} steps. Using complex analysis techniques like Mellin and
inverse Mellin transform, ~\cite{Fill:1} studied the asymptotic behavior of the first two moments of the algorithm {\em cost}. Moreover, they obtained the exact expression and asymptotic behavior of the distribution of \begin{math}H_n \end{math} and they have shown that a limit distribution for the centered algorithm {\em cost} \begin{math} H_n-\lfloor \log_2 n \rfloor \end{math} does not exist.  For a survey on Mellin transform, see
~\cite{Flajolet:1}.

\subsubsection*{The biased case}
If the splitting process is biased, \textit{i.e.} the probability that an item is sent into the left
subgroup is \begin{math}p\not =1/2 \end{math}, the algorithm is called {\em asymmetrical}. Studies on biased case become more rare. An asymmetric {\em leader election} algorithm was investigated by
~\cite{Janson:1} using complex analysis techniques. The asymptotic
behavior of the first two moments of the algorithm {\em cost}
\begin{math}H_n\end{math} is given in term of the sequence of their exact values \begin{math}\left(\mean(H_j)\right)_{j \in \N}\end{math}  computed numerically from two recurrence equations.

\medskip
\noindent
This implicit dependence is due to the asymmetry of the functional equation (\ref{eq:functeq}) obtained by Poissonization. The coefficient \begin{math}e^{-p\,x} \end{math} makes more complex the establishment of an iterative scheme such as in the context of a protocol for a multi-access broadcast channel (see ~\cite{Fayolle:1}).
Applying the Mellin transform to equation (\ref{eq:functeq}) without solving it yields this dependence.

\subsection{Related leader election algorithms}
\label{sec:var}

\subsubsection*{Leader election algorithm in network of fixed size}
Consider a simple algorithm for {\em leader election} algorithm in the
context of communication network; at each level, the probability
\begin{math}p\end{math} for a station to send its message depends on
  the number \begin{math}n\end{math} of stations remaining in the
    elimination process; \begin{math}p=1/n\end{math}. Expected run
    time is \begin{math}O(1)\end{math} but it is clear that
      is necessary to know the number of active stations in advance,
      or at least to estimate it. See ~\cite{Willard:1} for an estimation
    procedure in order of \begin{math}\log \log n+O(1/n)\end{math}. This variant of the basic {\em leader election} algorithm does not exhibit an oscillating behavior any more. In fact, the average algorithm cost is asymptotically equivalent to a some constant \begin{math}L\end{math}. For more details, see ~\cite{Lavault:2}.

\subsubsection*{LZ'\begin{math}77\end{math} data compression Scheme}
Consider a variant of the {\em leader election} algorithm by introducing a moderator who determines the elimination process; each of participants and the moderator throws independently a coin and only those who obtain the same result as the moderator continue the process.
See ~\cite{Ward:1} for the biased case, ~\cite{Prodinger:1} for the
unbiased one. Let  \begin{math}M_n\end{math} the number of participants remaining in the last nontrivial round from an initial group of \begin{math}n\end{math} items. It is asymptotically equivalent to the multiplicity of phrases in the LZ'\begin{math}77\end{math} data compression scheme.

\subsection{Overview}
In a previous paper on {\em splitting} algorithms, ~\cite{Mohamed:1} proposed a direct approach based on a probabilistic reformulation of a basic functional equation associated to such algorithms. The purpose of this work is to apply the techniques used by ~\cite{Mohamed:1} to analyze an additive quantity in the context of an incomplete tree structure. In Section \ref{sec:ave}, a similar series formula for the average {\em cost} \begin{math}E(H_n)\end{math} is given by Proposition \ref{cost-exact}. The asymptotic behavior of the algorithm is studied in Section \ref{sec:asym} and reformulated on the behavior of some stopping time \begin{math}\tau \end{math}. Theorem \ref{theo-asymp} presents a new representation of the asymptotic oscillations of the algorithm. In Section  \ref{sec:dist}, the distribution of the algorithm {\em cost} is investigated. Using the binary
decomposition  of the interval \begin{math}[0,1]\end{math}, the exact expression of the distribution of \begin{math}H_n\end{math} is established.  Proposition \ref{prop-dist} is a slight variation of the asymptotic formula given by  ~\cite{Janson:1} for the distribution of the algorithm {\em cost} \begin{math} H_n \end{math} in the biased case.

\subsection{Notations}
\label{sec:notations}
Throughout this paper, \begin{math}(t_n)_{n\geq 1}\end{math} is a non
decreasing random variables sequence such that
\begin{itemize}
\item \begin{math}t_1\end{math} follows an exponential distribution with parameter \begin{math}1\end{math},
\item \begin{math}(t_{n+1}-t_n)\end{math} is a sequence
of {\em i.i.d.} random variables exponentially distributed with
parameter \begin{math}1\end{math}.
\end{itemize}
For \begin{math}x \geq 0\end{math}, let \begin{math}{\cal N}_x\end{math} be the number of \begin{math}t_n\end{math} in the interval \begin{math}[0,x]\end{math}. It is a r. v. with {\em Poisson} distribution  .

\section{Average Cost of The Algorithm}
\label{sec:ave}

\subsection{Algorithm cost}
The algorithm {\em cost} is the number of steps needed to find a {\em leader}, or equivalently the {\em height} of the associated tree. Denote
by \begin{math}H_n\end{math} this quantity when the size of the initial group of items
is \begin{math}n\end{math}, then, for \begin{math}n \geq 2\end{math}, this random variable verifies a recurrence relation;
\[
 H_n \stackrel{{dist.}}{=} 1+H_{1,S_n}\, {{\ind}_{\{S_n \not = 0\}}}+H_{2,n} \, 
{{\ind}_{\{S_n =0\}}},
\]
with the boundary conditions  \begin{math}H_0 =H_1=0\end{math},
where \begin{math}(B_i (p))_{1 \le i \le n}\end{math} are \begin{math}n\end{math} independent {\em
Bernoulli} variables of parameter \begin{math}p\end{math},
\[
S_n=\sum_{i=1}^n B_i(p),
\]
for \begin{math}(m,n) \in \N^2\end{math}, \begin{math}H_{1,m}\end{math} and \begin{math}H_{2,n}\end{math} are independent and, for \begin{math}i=1,2\end{math}, the variable \begin{math}H_{i,m}\end{math} has the same distribution as 
\begin{math}H_m\end{math}. So, for \begin{math}n \geq 0\end{math}, the recurrence equation for the sequence \begin{math}(H_n)\end{math} can be rewritten
\begin{equation}
 H_n \stackrel{{dist.}}{=} 1+H_{S_n}+H_n \, {\ind}_{(S_n =0)}-{\ind}_{\{n \le 1\}}.\label{eq:reclimit}
\end{equation}

\subsection{Poissonization}

Consider the {\em Poisson} model, \textit{i.e.} the size of the initial group of items is random following a {\em Poisson} process ${\cal N}_x$ of intensity \begin{math}1 \end{math} on the interval \begin{math}[0,x] \end{math}. The following proposition gives a useful representation of the {\em Poisson 
transform} of the average {\em cost} of the algorithm.
\begin{proposition}
For \begin{math}x>0\end{math},
\[
\mean(H_{{\cal N}_x})=\mean\left(\sum_{i=0}^{+ \infty} \frac{1}{\pi_i}\, 
{\ind}_{\{t_1>x \pi_i \,;\, t_2 \le x (\alpha_i+\pi_i)\}}\right),
\]
where \begin{math}(A_j,B_j)\end{math} is a sequence of i.i.d. realizations of a couple of random variable \begin{math}(A,B)\end{math} with distribution
\[
\P(A=p,B=0)=p,~\P(A=q,B=p)=q,
\]
\begin{math}\pi_0=1,~\alpha_0=0\end{math} and, for \begin{math}i \geq 1\end{math},
\[
\pi_i=\prod_{j=0}^{i-1}A_j,~\alpha_i= \sum_{j=0}^{i-1}\pi_j \,B_j.
\]
\end{proposition}

\begin{proof}
Let \begin{math}h\end{math} the {\em Poisson transform} of the average {\em cost} (see Definition \ref{def}). Then, the recurrence equation \begin{math}(\ref{eq:reclimit})\end{math} for the sequence \begin{math}(H_n)_{n\geq 0}\end{math} becomes
\[
h(x)=h(px)+h(qx)\, e^{-px}+1-(1+x)e^{-x}.
\]
Following the approach of ~\cite{Mohamed:1}, direct iteration becomes possible using a probabilistic formulation of the last equation as below
\begin{equation}
h(x)=\mean \left(\frac{h(A x)}{A}\, e^{-Bx}\right)+f(x), \label{eq:probaeq}
\end{equation}
where \begin{math} f(x)=1-(1+x)e^{-x} \end{math} and \begin{math}(A,B)\end{math} is couple of random variables with distribution 
\[
\P(A=p,B=0)=p,~\P(A=q,B=p)=q.
\]
Let the sequence of {\em i.i.d} realizations \begin{math}(A_i,B_i)_{i 
\in
\N}\end{math} of the couple
of random variables  \begin{math}(A,B)\end{math}. We introduce some notations; for  \begin{math}x \geq 0\end{math}, \begin{math}X_0=x,~Y_0=0\end{math}, and for \begin{math}n \in \N \end{math},
\[
X_{n+1}=A_n \, X_n,~Y_{n+1}=B_n \, X_n.
\]
By iterations of equation (\ref{eq:probaeq}), one gets at the \begin{math}(n+1)^{th}\end{math} stage
\[
h(x)=\mean \left(\frac{h(X_{n+1})}{\prod_{i=0}^n A_i}\, e^{-\sum_{i=0}^{n+1}Y_i}\right)
+\mean \left( \sum_{i=0}^{n}e^{-\sum_{j=0}^{i}Y_j}\, \frac{f(X_i)}{\prod_{j=0}^{i-1}A_j}\right).
\]
Since \begin{math}h'(0)=0\end{math} and, almost surely,
\begin{math}
\lim_{n \rightarrow + \infty} X_{n+1}=0\end{math}, then, one obtains
\[
h(x)=\mean \left(\sum_{i=0}^{+
\infty}\frac{1}{\pi_i}\left(1-(1+\pi_i)e^{-\pi_i x}\right) e^{-\alpha_i x}\right),
\]
where \begin{math}\pi_0=1,~\alpha_0=0\end{math} and, for \begin{math}i \geq 1\end{math},
\[
\pi_i=\prod_{j=0}^{i-1}A_j,~\alpha_i=\sum_{j=0}^{i-1}\pi_j \,B_j.
\]
As the sequences \begin{math}(\alpha_i)\end{math} and \begin{math}(\alpha_i+\pi_i)\end{math} are, almost
surely, in the interval \begin{math}[0,1]\end{math}, the function \begin{math}h\end{math} can be
represented as follows
\begin{equation}
h(x)=\mean \left (\sum_{i=0}^{+ \infty}\frac{1}{\pi_i}\,
{\ind}_{\{t_1>\alpha_i x \, ; \, t_2<(\alpha_i+\pi_i)x\}}\right ).\label{eq:poissonaverage}
\end{equation}
The proposition has been proved.
\end{proof}

\medskip
\noindent
>From now on, throughout the paper, we conserve the notations
introduced in this proof.

\subsection{de-Poissonization}
The next step is the probabilistic de-Poissonization of \begin{math}(\ref{eq:poissonaverage})\end{math} following the method of  ~\cite{robert:1} to obtain the expression of the average {\em cost} \begin{math}\mean(H_n)\end{math}.
\begin{proposition}[Probabilistic representation of the average cost]\label{cost-exact}
For \begin{math}n \geq 2\end{math}, 
\[
\mean(H_n)=\mean \left 
(\sum_{i=0}^{\tau(U_{1,n},U_{2,n})-1}\frac{1}{\pi_i}\right ),
\]
where, for \begin{math}0<x<y<1, ~\tau(x,y)=\min \left (\nu(x);\mu(y) \right )\end{math} with
\begin{eqnarray*}
\nu(x)&=&\inf \left \{i\geq 1: \alpha_i >x\right \}, \nonumber \\
\mu(y)&=&\inf \left \{i\geq 1: \alpha_i+\pi_i <y\right \}, \nonumber
\end{eqnarray*}
and \begin{math}U_{i,n}\end{math} is the \begin{math}i\end{math}th smallest variables of \begin{math}n\end{math} independent,
uniformly distributed random variables on \begin{math}[0,1]\end{math} independent of the 
sequence
\begin{math}\left (A_j,B_j \right )_{j \geq 0}\end{math}.
\end{proposition}
\begin{proof}
For \begin{math}x>0\end{math}, by decomposing with respect to the number of points of
the Poisson process \begin{math}({\cal N}_x)\end{math} in the interval \begin{math}[0,x]\end{math}, one gets,
for \begin{math}0<a<b<1\end{math},
\[
\P(t_1>ax \, ,\,  t_2<bx)=\sum_{n=2}^{+ \infty}\P(t_1>ax \, , \, t_2<bx | {\cal N}_x=n) 
\P({\cal N}_x=n).
\]
For \begin{math}n \geq 2\end{math}, conditionally on the event \begin{math}\{{\cal N}_x=n\}\end{math}, the
couple of variables \begin{math}(t_1,t_2)\end{math} has the same distribution as the
couple \begin{math}(xU_{1,n}, xU_{2,n})\end{math} of the two smallest random variables
of \begin{math}n\end{math} uniformly distributed random variables on \begin{math}[0,x]\end{math}. So, we
get the identity
\[
\P(t_1>ax , \, t_2<bx)=\mean \left(\sum_{n=2}^{+ \infty}{\ind}_{\{U_{1,n}>a , \, 
U_{2,n}<b\}}\frac{x^n}{n!}e^{-x}\right ).
\]
Due to the independence of the sequence \begin{math}(A_i,B_i)\end{math} and \begin{math}(t_1,t_2)\end{math}, and using the Fubini's Theorem, one gets
\[
\mean(H_{{\cal N}_x})=\sum_{n=2}^{+ \infty}\left (\mean \left (\sum_{i=0}^{+ 
\infty}\frac{1}{\pi_i}\, {\ind}_{\{U_{1,n}>\pi_i \, , \, 
U_{2,n}<(\alpha_i+\pi_i)\}}\right )\right )\frac{x^n}{n!}e^{-x}.
\]
The identification of the representation of the {\em Poisson transform} (see Definition \ref{def}) 
\begin{math}\mean(H_{{\cal N}_x})\end{math} and the last identity gives the following
formula for \begin{math}n\geq 2\end{math}
\[
\mean(H_n)=\mean \left (\sum_{i=0}^{+ \infty}\frac{1}{\pi_i}\,
{\ind}_{\{U_{1,n}>\pi_i \, , \, U_{2,n}<(\alpha_i+\pi_i)\}}\right ).
\]
Since, almost surely, the sequence \begin{math}(\alpha_i)_{i \geq 0}\end{math} is
increasing to a random variable \begin{math}\alpha \in [0,1]\end{math} and the
sequence \begin{math}(\alpha_i+\pi_i)_{i \geq 0}\end{math} is decreasing to the same
random variable, the following equality holds
\[
\{i \geq 0\, : U_{1,n}>\pi_i \, , \,
U_{2,n}<\alpha_i+\pi_i\}=[0,\tau(U_{1,n},U_{2,n})-1],
\]
where the hitting time \begin{math}\tau \end{math} is defined as above.
\end{proof}

\section{Asymptotic Analysis of The Average Cost}
\label{sec:asym}
\subsection{Two random sequences and one hitting time}
\label{sec:rand}
 It is clear that the key of the analysis of the asymptotic behavior of the algorithm is the hitting time \begin{math}\tau\end{math} written on the two random sequences  \begin{math}(\alpha_i)_{i \geq 0}\end{math} and \begin{math}(\alpha_i+\pi_i)_{i \geq 0}\end{math}. Let \begin{math}(\gamma_i)\end{math} the sequence of random variables defined by
\[
\left (\gamma_i\right )_{i \in \N}=\{j \geq 0\, :\, B_j=p\}.
\]
These are the moments of jump of the sequence \begin{math}(\alpha_i)_{i \geq
  0}\end{math} and conversely the instants of stopping for the other one,
\begin{math}(\pi_i+\alpha_i)_{i \geq 0}\end{math}. It is clear that these moments can be
  recursively defined as below: \begin{math}\gamma_0=G_0\end{math} and \begin{math}\gamma_{n+1}=1+\gamma_n +G_{n+1}\end{math}, where \begin{math}\left(G_n \right)_{n \geq 0}\end{math} is a sequence of {\em i.i.d} r.
v. with a geometric distribution \begin{math}Geo (q)\end{math}
\[
\P(Geo(q)=k)=q~p^k.
\]
So, it is easy to see that
\[
\nu(x) \in \{1+\gamma_i \, : \, i \in \N\}, ~\mu(y) \not \in \{1+\gamma_i \, : \, i \in \N\}.
\]
Using a discussion on the position of the hitting time \begin{math} \tau \end{math} in comparison with the sequence \begin{math}\gamma\end{math}, we establish the following lemma which will be proved in the Appendix \ref{sec:app}.
\begin{lemma}\label{eq:lem}
\begin{eqnarray*}
\mean \left (\sum_{i=0}^{\tau(x,y)-1}\frac{1}{\pi_i} \right )&=&\lceil
\log_p(y) \rceil + \left(\lceil
  \log_p(\rho(\log_p(y))y) \rceil-\lfloor \log_p(y)\rfloor  \right){\ind}_{\Omega(x,y)}\nonumber \\
&+&\mean \left ( \sum_{i=1+\lceil
  \log_p(\rho(\log_p(y))y) \rceil}^{\tau(x,y)-1}\frac{1}{\pi_i} \,
{\ind}_{\{\gamma_0=\lfloor \log_p(y) \rfloor ; \gamma_1=\lceil
  \log_p(\rho(\log_p(y))y) \rceil \}} \right ){\ind}_{\Omega(x,y)}.\nonumber
\end{eqnarray*}
where  \begin{math} \Omega(x,y)=\{(x,y) \in (]0,1[)^2 \, : \,\lceil \log_p(y)
\rceil=\lceil \log_p(x) \rceil \}\end{math}  and \begin{math} \rho \end{math} is a periodic function with magnitude \begin{math}1\end{math} defined for \begin{math} z>0\end{math} by
\[
\rho(z)=\frac{1-p^{1-\{z\}}}{1-p}~,\mbox{  \begin{math}\{z\}=z-\lfloor z \rfloor \end{math} is the fractional part of \begin{math} z \end{math}}.
\]
\end{lemma}

\subsection{Asymptotic fluctuations phenomena} 

\begin{theorem}[Asymptotic behavior of the average cost]\label{theo-asymp}
The average cost \begin{math}\mean(H_n)\end{math} admits the following asymptotic formula
\[
\mean(H_n)=-\log_p(n)+\mean \left (\lceil \log_p (t_2) \right\rceil )+F(\log_p(n))
+\cal{R}(n),
\]
where \begin{math}F\end{math} is a periodic function defined for all \begin{math}z>0\end{math} by
\begin{equation}\label{eq:F}
F(z)=\int_0 ^\infty y(1-p^{1-\{\log_p y-z\}}) 
\left (\lceil \log_p(\frac{1-p^{1-\{\log_p y-z\}}}{1-p})+\log_p y-z
\rceil
-\lfloor \log_p y -z \rfloor \right ) e^{-y} dy,
\end{equation}
\begin{math} \Omega_{n}=\Omega(U_{1,n},U_{2,n}) \end{math}  and \begin{math}\cal{R}(n)\end{math} is a rest discussed in Section \begin{math}\ref{sec:rest}\end{math}, defined by
\begin{equation}\label{eq:rest}
{\cal{R}}(n)=\mean \left(
(\sum_{i=1+\lceil
  \log_p(\rho(\log_p(U_{2,n}))U_{2,n}) \rceil}^{\tau(U_{1,n},U_{2,n})-1}\frac{1}{\pi_i}) {\ind}_{\{\gamma_0 = 
\lfloor
\log_p(U_{2,n}) \rfloor ; \gamma_1=\lceil \log_p(\rho(U_{2,n})U_{2,n}) \rceil\}}\, {\ind}_{\Omega_n}\right) .
\end{equation}
\end{theorem}

\begin{proof}
Using Lemma \begin{math}\ref{eq:lem}\end{math}, one gets
\begin{eqnarray*}
 \mean(H_n)&=&\mean \left (\lceil \log_p(U_{2,n}) \rceil \right )+\mean \left ((\lceil \log_p(\rho(\log_p(U_{2,n})) U_{2,n}) 
\rceil-\lfloor \log_p( U_{2,n}) \rfloor ) 
{\ind}_{\Omega_{n}} \right )
\nonumber \\
&+&\mean \left(
(\sum_{i=1+\lceil
  \log_p(\rho(\log_p(y))y) \rceil}^{\tau(U_{1,n},U_{2,n})-1}\frac{1}{\pi_i}){\ind}_{\{\gamma_0 = 
\lfloor
\log_p(U_{2,n}) \rfloor ; \gamma_1=\lceil \log_p(\rho(U_{2,n})U_{2,n}) \rceil\}}\, {\ind}_{\Omega_n} \right).
\nonumber
\end{eqnarray*}
The only not neglect
terms are
\[
\cal{T}_1(n)=\mean \left (\lceil \log_p(U_{2,n}) \rceil \right ) ~\mbox{ and }~
\cal{T}_2(n)=\mean \left( (\lceil \log_p(\rho(\log_p(U_{2,n})) U_{2,n}) 
\rceil-\lfloor \log_p( U_{2,n}) \rfloor ) 
{\ind}_{\Omega_{n}} \right).  
\]
As \begin{math}n\end{math} goes to infinity, \begin{math}n U_{2,n}\end{math} converges in distribution to a random variable \begin{math}t_2\end{math} which is a sum of two {\em i.i.d.} exponential random variables with parameter \begin{math}1\end{math}.Then, the first
term satisfies
\[
\cal{T}_1(n)=\mean (\lceil \log_p(t_2)- \log_p(n) \rceil )+O(\frac{1}{n}).
\]
Let \begin{math}\cal{D}\end{math}, function of \begin{math}-\log_p(n)\end{math}, the difference
\[
\cal{D}(-\log_p(n))=\mean (\lceil \log_p(t_2)- \log_p(n) \rceil )-\left(\mean (\lceil \log_p(t_2) \rceil )- \log_p(n)\right)
\]
It is easy to check that \begin{math}\cal{D}(z)=\cal{D}(\{z\})-\lfloor z \rfloor\end{math}, then \begin{math}
\lim_{n \rightarrow +\infty}n~\cal{D}(-\log_p n)=\lim_{z \rightarrow +\infty}p^{-z}~\cal{D}(z)=0 \end{math}, and one gets 
\[
\cal{T}_1(n)=- \log_p(n)+\mean (\lceil \log_p(t_2) \rceil )+O(\frac{1}{n}).
\]
The last term \begin{math}\cal{T}_2(n)\end{math} is asymptotically equivalent to \begin{math}F(\log_p(n))\end{math} where \begin{math}F\end{math} is defined by ~\ref{eq:F}.
In fact
\[
|F(\log_p(n))-\cal{T}_2(n)|\le \int_0^n \left|(1-\frac{y}{n})^{n-2}-e^{-y}\right|dy
+\int_n ^\infty \log_p(\rho(\log_p(y/n)))~y~ e^{-y} dy
+\frac{1}{n}F(\log_p(n))+2 e^{-n}
\]
Observe that
\[
\int_n ^\infty \log_p(\rho(\log_p(y/n)))~y~ e^{-y} dy=n^2\int_1^{\infty}\log_p(\rho(\log_py))y~e^{-ny}dy. 
\]
By decomposition on the sequence of intervals \begin{math}\left([p^{k+1},p^{k}]\right)\end{math}, the last integral is dominated by a geometric  sum and the following inequality holds for \begin{math}n>2\end{math}
\[
\int_1 ^\infty \log_p(\rho(\log_p(y)))~y~ e^{-ny} dy \le \frac{p^{n-2}}{1-p}.
\]
Then,
\[
F(\log_p(n))-\mean \left( (\lceil \log_p(\rho(\log_p(U_{2,n})) U_{2,n}) 
\rceil-\lfloor \log_p( U_{2,n}) \rfloor ) 
{\ind}_{\Omega_{n}} \right)=O(\frac{1}{n}).
\]
This ends the proof.
\end{proof}

\subsection{Estimation of the rest}
\label{sec:rest}
The final step is to estimate the rest \begin{math}{\cal{R}}(n)\end{math} defined by \begin{math}(\ref{eq:rest})\end{math}. For \begin{math}x,y \in [0,1]\end{math}, \begin{math}K>k>0\end{math}
\[
\mean \left ( \sum_{i=K}^{\tau(x,y)-1}\frac{1}{\pi_i} \,
{\ind}_{\{\gamma_0=k ; \gamma_1=K\}} \right )\le  (1-\delta)^K \sqrt{\mean \left((\frac{1}{\delta^2})^{\tau(x,y)}\right)},
\]
where \begin{math}\delta=\min(p,q)\end{math}. The following result is admitted.
\begin{conjecture}\label{eq:conj}
The hitting time \begin{math}\tau\end{math} satisfies
\[
\sup_{x\in [0,1]} \mean \left((\frac{1}{\delta^2})^{\tau(x,x)}\right)<\infty.
\]
\end{conjecture}

\begin{figure}[htbp]
  \begin{center}
    \includegraphics[height=7cm]{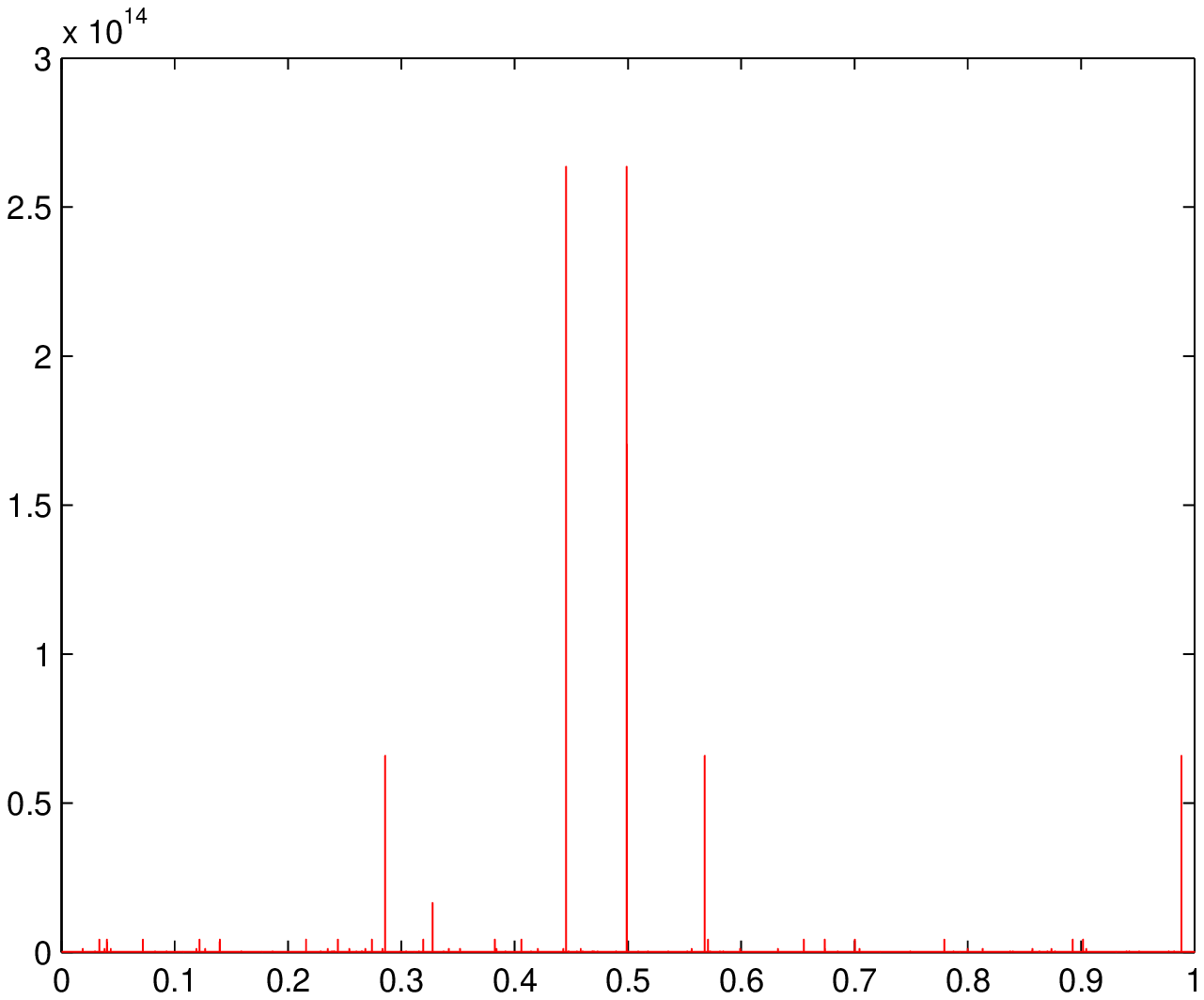}
    \caption{Unbiased case: simulations of $x: \rightarrow \mean \left(4^{\tau(x,x)}\right)$.}
    \label{fig:unbiasedTAU}
  \end{center}
\end{figure}

\begin{figure}[htbp]
  \begin{center} 
    \subfigure[$p=0.2,~\delta=p$\label{0.2}]{\includegraphics[height=5cm]{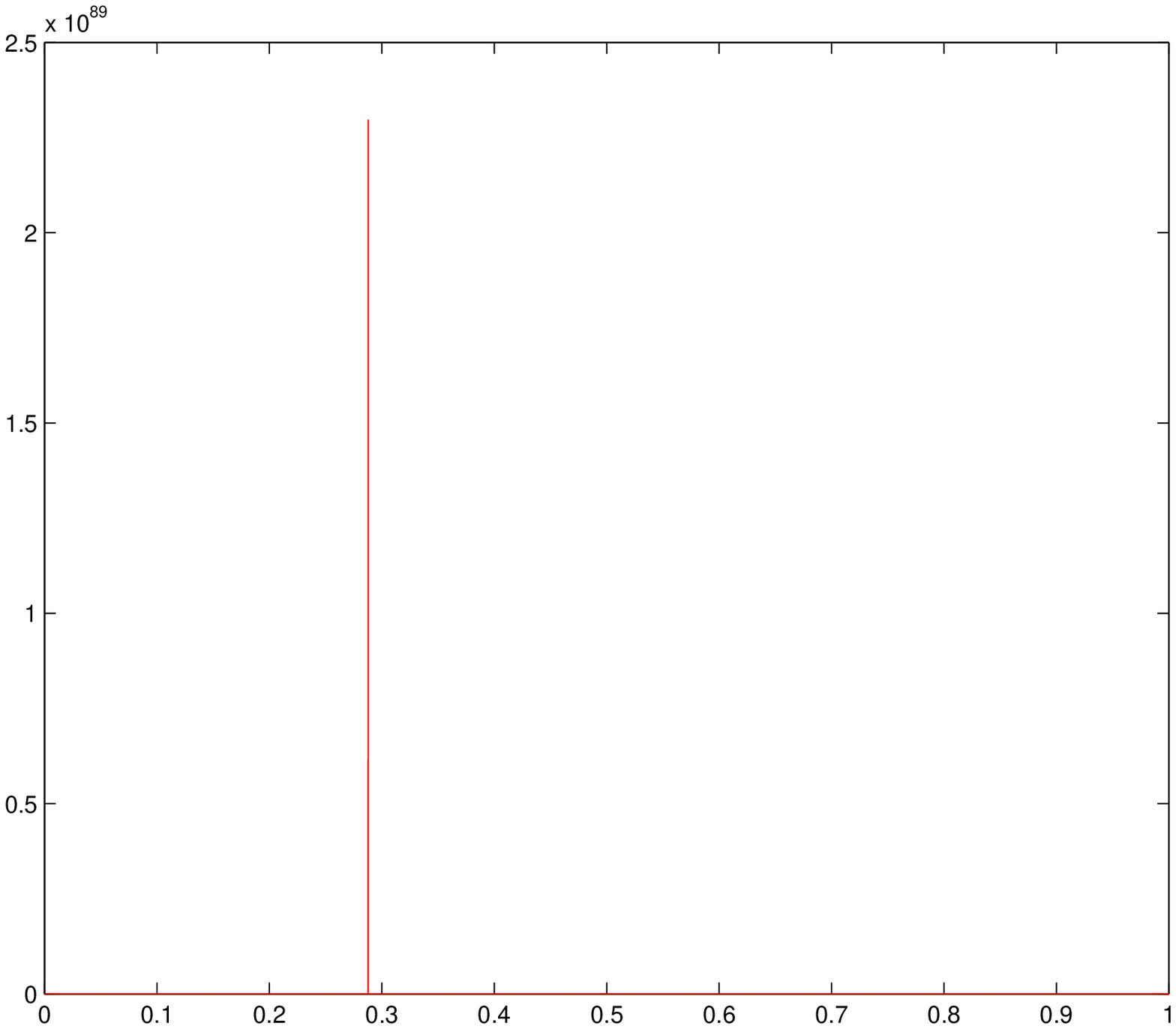}}
    \hfil
    \subfigure[$p=0.8,~\delta=1-p$\label{0.8}]{\includegraphics[height=5cm]{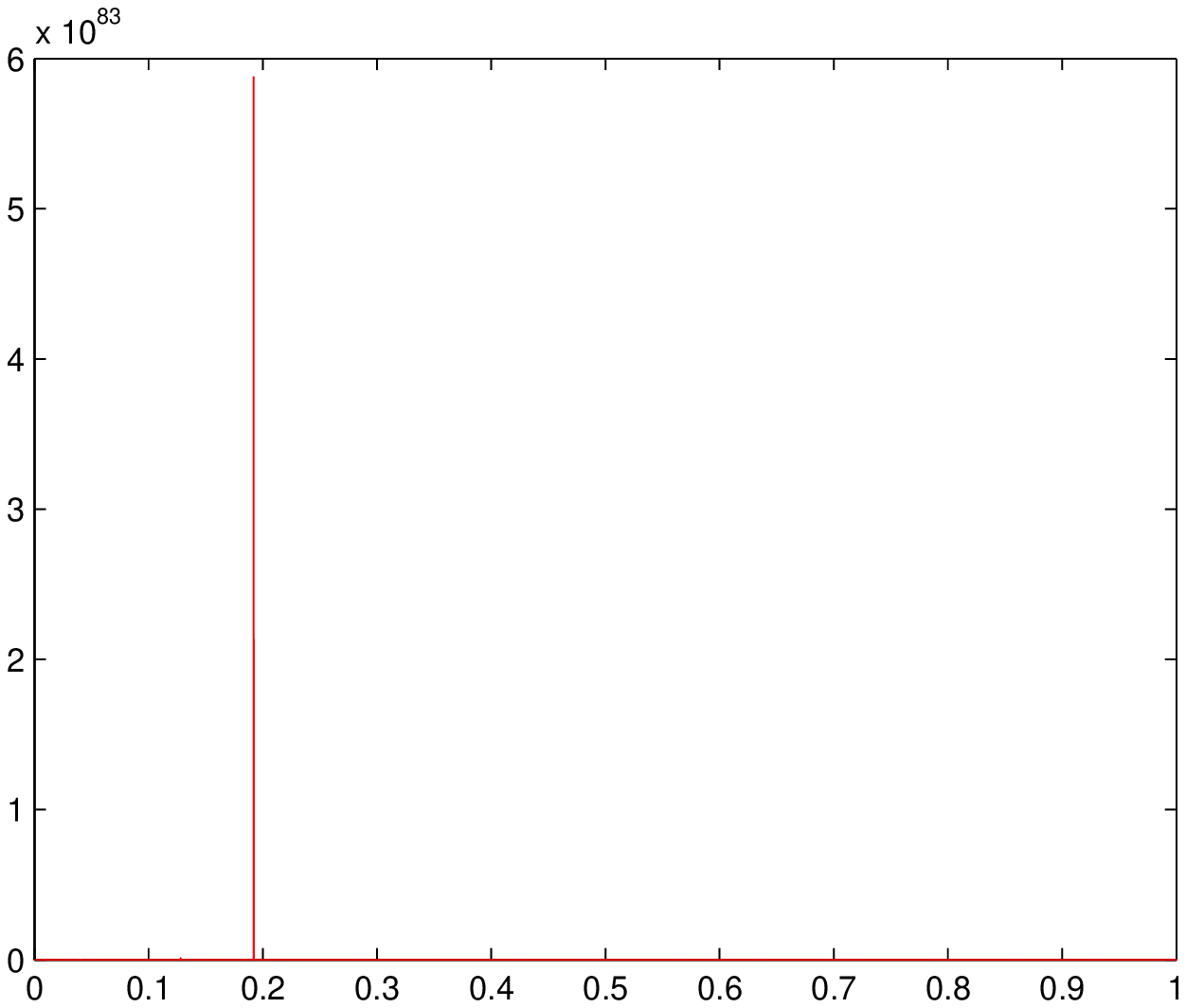}}
    \caption{Biased case: simulations of $x: \rightarrow \mean \left(\frac{1}{\delta^2}^{\tau(x,x)}\right)$ .}
    \label{fig:biasedTAU}
  \end{center}
\end{figure}

\begin{remark}
Conjecture \ref{eq:conj} is an intuitive restriction on the exponential moment of the hitting time \begin{math} \tau \end{math}. It is supported by some simulations (Fig.\ref{fig:unbiasedTAU},\ref{fig:biasedTAU}) of $x: \rightarrow \mean \left((\frac{1}{\delta^2})^{\tau(x,x)}\right)$ using Monte-Carlo techniques. Observe that, for the unbiased case (Fig.\ref{fig:unbiasedTAU}), the maximum corresponds to numerical values of \begin{math} x\end{math} around \begin{math}0.5\end{math} which is, on average, the limit \begin{math} \alpha \end{math} of the two random sequences \begin{math} (\alpha_i)\end{math} and \begin{math} (\alpha_i+\pi_i)\end{math}. This maximum is of order of \begin{math} 10^{14}\end{math}, which is reasonable since it implies that
\[
\mean(\tau)\le 14~\log_4(10)\approx 23.25
\]
For the biased one (Fig.\ref{fig:biasedTAU}),  since \begin{math} \delta=0.2\end{math}, a maximum of the order of \begin{math}10^{80}\end{math} is acceptable; \begin{math}\mean(\tau)\le 57.22 \end{math}.
\end{remark}
%*******************************

%**********************************
%***********************************

\medskip
\noindent
Since, for \begin{math} 0\le x<y \le1\end{math}, \begin{math}
\tau(x,y)\le \max(\tau(x,x),~\tau(y,y))
\end{math}, then, using Conjecture \ref{eq:conj}, we obtain
\[
{\cal{R}}(n)\le C~\mean \left((1-\delta)^{\lceil \log_p(\rho(U_{2,n})U_{2,n}) \rceil}\right),
\]
where
\begin{math}
C=\sup_{ x\in[0,1]} \sqrt{\mean \left((1/\delta^2)^{\tau(x,x)}\right)}.
\end{math}
Using the same method as for the function \begin{math}F\end{math}, one gets
\[
\mean \left((1-\delta)^{\lceil \log_p(\rho(U_{2,n})U_{2,n}) \rceil}\right) \le \mean \left((1-\delta)^{ \lceil \log_p(U_{2,n}) \rceil}\right)\sim \left(\frac{1}{n}\right)^{\log_p(1-\delta)}.
\]
This gives
\[
{\cal{R}}(n)=O(\frac{1}{n^{\log_p(1-\delta)}}).
\]
Conclusion
\[
 \mean(H_n) =-\log_p(n)+
\mean \left( \lfloor \log_p(t_{2}) \rfloor \right) +F(\log_p(n))+O(\frac{1}{n^{\log_p(1-\delta)}}).
\]

\section{Algorithm Cost Distribution}
\label{sec:dist}
It is more appropriate to use these notations \begin{math}p_0=p,\,p_1=q\end{math}, to define recursively the sequence of intervals \begin{math}\left (I_k^n \right
)\end{math} associated to the binary decomposition of the interval \begin{math}[0,1]\end{math} in the base \begin{math}(p_0,\,p_1)\end{math}
\[
\left \{ \begin{array}{ll}
I_0 ^0 &=[0,1]\\
I_k ^{n+1} &=\left (I_{k-1}^{n+1}\right )_+ +\, p_{k-2 \lfloor k/2 
\rfloor}\, I_{\lfloor k/2 \rfloor}^n,
\end{array}
\right.
\]
where \begin{math}\left (I\right )_+\end{math} denotes the right extremity of the
interval \begin{math}I\end{math}. Let \begin{math}|I|\end{math} the length of the interval \begin{math}I\end{math}, then
\[
\left (I_k ^n \right )_+=\sum_{i=0}^{k}|I_i ^n|.
\]
Let \begin{math}n \in \N\end{math} and \begin{math}0 \le k \le 2^{n+1}-1\end{math}. Consider the binary
decomposition of \begin{math}k\end{math} at the stage \begin{math}n\end{math}
\[
k=a_0+a_1 2+\ldots+a_n 2^n,~\mbox{ for \begin{math}0 \le i \le n\end{math}, \begin{math}~a_i \in \{0,1\}\end{math}.}
\]
Then, the length of the
interval \begin{math}I_k ^{n+1}\end{math} is 
\[
|I_k ^{n+1}|=\prod_{i=0}^{n}p_{a_i}.
\]
For \begin{math}k \in \N\end{math}, \begin{math}x>0\end{math}, one gets the following identity
\[
\{H_{{\cal N}_x}>k\}=\{\exists~ 0\le i <2^k \, :\,{\cal N}(x I_0 ^k)=\ldots={\cal N}(x 
I_{i-1} ^k)=0\, , \, {\cal N}(x I_{i} ^k)\geq 2\}.
\]
So
\begin{equation}
\P(H_{{\cal N}_x}\le k)=e^{-x}+x \, \sum_{i=0}^{2^k-1}|I_i ^k|e^{-\left 
(I_i ^k \right )_+}.\label{eq:hdensite}
\end{equation}
Let us define the sequence of probability measures \begin{math}\left (\mu_k \right 
)\end{math} by
\[
\mu_k(t)=\sum_{i=0}^{2^k-1}|I_i ^k|\, \delta_{\left (I_i ^k \right 
)_+}(t).
\]
Then, equation \begin{math}(\ref{eq:hdensite})\end{math} can be rewritten as
\begin{equation}
\P(H_{{\cal N}_x}\le k)=e^{-x}+x \int_0 ^1 e^{-xt}\, d\mu_k(t)\label{eq:poissonDensite}.
\end{equation}
Using a probabilistic de-Poissonization of equation \begin{math}(\ref{eq:poissonDensite})\end{math} as done for  Proposition \ref{cost-exact}, we obtain the exact distribution of \begin{math}H_n\end{math}.
\begin{proposition}\label{prop-dist}
For \begin{math}n\geq 2\end{math},
\[
\P({H_n}\le k)=n \int_0 ^1 (1-t)^{n-1}\, d\mu_k(t),
\]
where the probability measure \begin{math}\mu_k\end{math} is described as above.
\end{proposition}
Using this identity
\[
1-n(1-t)^{n-1}=1-nt\, (1-t)^{n-1}-n(1-t)^n,
\]
the following result is immediate.
\begin{corollary}
For \begin{math}k \in \N\end{math},
\[
\P(H_n >k) \sim \int_0 ^1 \P(U_{2,n}<t)\, d\mu_k (t) \mbox{ , as \begin{math}n\end{math} 
goes to infinity}.
\]
where \begin{math}U_{2,n}\end{math} is the second smallest random variable of \begin{math}n\end{math} uniformly 
distributed random variables on \begin{math}[0.1]\end{math}
\end{corollary}

\section{Appendix}
\label{sec:app}
We present the proof of Lemma \begin{math}\ref{eq:lem}\end{math}. Recall the sequence of random variables \begin{math}(\gamma_i)\end{math} defined by
\[
\left (\gamma_i\right )_{i \in \N}=\{j \geq 0\, :\, B_j=p\}.
\]
\begin{proof}[of Lemma \begin{math}\ref{eq:lem}\end{math}]
First, note that
\begin{eqnarray*}
(\nu(x) \geq 2+\gamma_0) & \Leftrightarrow& (\gamma_0 \geq \lfloor
\log_p(x) \rfloor) \nonumber \\
(\mu(y) \leq \gamma_0) & \Leftrightarrow& (\gamma_0 \geq \lceil
\log_p(y) \rceil). \nonumber
\end{eqnarray*}
Denote by \begin{math}\Omega_0\end{math} the following set
\[
\Omega_0=\Omega(x,y):=\{(x,y) \in (]0,1[)^2 \, : \,\lceil \log_p(y)
\rceil=\lceil \log_p(x) \rceil \}.
\]
By decomposing the function \begin{math}\Phi\end{math} with respect to \begin{math}\Omega_0\end{math}, one gets
this formula
\begin{eqnarray*}
\mean \left (\sum_{i=0}^{\tau(x,y)-1}\frac{1}{\pi_i}\right )&=&\mean 
\left (\sum_{i=0}^{\gamma_0}\frac{1}{p^i}{\ind}_{\{\gamma_0<\lfloor \log_p(y) 
\rfloor \}}\right ) 
+\P \left( \gamma_0=\lfloor \log_p(y) \rfloor \right)
\sum_{i=0}^{\lfloor \log_p(y) \rfloor}\frac{1}{p^i} 
+\mean \left
(\sum_{i=0}^{\mu(y)-1}\frac{1}{p^i}{\ind}_{\{\gamma_0\geq \lceil
\log_p(y) \rceil \}}\right )\nonumber \\
&+&\mean \left
((\sum_{i=\lceil \log_p(y) \rceil}^{\tau(x,y)-1}\frac{1}{\pi^i}){\ind}_{\{\gamma_0 = \lfloor
\log_p(y) \rfloor \}}\right ){\ind}_{\Omega_0}.\nonumber
\end{eqnarray*}
Since 
\[
\mu(y)|(\gamma_0\geq \lceil \log_p(y) \rceil)=\inf\{i \geq 1, p^i< y 
\}= \lceil \log_p(y) \rceil,
\]
then, by simple calculations, one gets
\[
 \mean \left (\sum_{i=0}^{\tau(x,y)-1}\frac{1}{\pi_i}\right )=\lceil
\log_p(y) \rceil 
+\mean \left
((\sum_{i=\lceil \log_p(y) \rceil }^{\tau(x,y)-1}\frac{1}{\pi^i}){\ind}_{\{\gamma_0 = \lfloor
\log_p(y) \rfloor \}}\right ){\ind}_{\Omega_0}.
\]
A second discussion on \begin{math}\gamma_1\end{math} implies that, on the set \begin{math}(\gamma_0 = 
\lfloor
log_p(y) \rfloor, \Omega_0)\end{math},
\begin{eqnarray*}
(\nu(x) \geq 2+\gamma_1) & \Leftrightarrow& (\gamma_1 \geq \lceil
\log_p(\rho(\log_p(x)) x) \rceil) \nonumber \\
(\mu(y) \leq \gamma_1) & \Leftrightarrow& (\gamma_1 \geq 1+\lceil
\log_p(\rho(\log_p(y)) y) \rceil), \nonumber
\end{eqnarray*}
where \begin{math} \rho \end{math} is a periodic function with magnitude \begin{math}1\end{math} defined by \begin{math}
\rho(z)=(1-p^{1-\{z\}})/(1-p)
\end{math}. Moreover, \begin{math}\rho\end{math} is decreasing on \begin{math}[0,1[\end{math}, so on the set \begin{math}\Omega_0\end{math}, 
\[
\rho(\log_p(x))x<\rho(\log_p(y))y.
\]
Let \begin{math} \Omega_1=\Omega(\rho(\log_p(x))x,\rho(\log_p(y))y) \end{math}.
Then
\begin{eqnarray*}
 \mean \left (\sum_{i=0}^{\tau(x,y)-1}\frac{1}{\pi_i}\right )&=&\lceil 
\log_p(y) \rceil 
+\left(\lceil
  \log_p(\rho(\log_p(y))y) \rceil-\lfloor \log_p(y)\rfloor  \right){\ind}_{\Omega(x,y)}\nonumber \\
&+&\mean \left ( \sum_{i=1+\lceil
  \log_p(\rho(\log_p(y))y) \rceil}^{\tau(x,y)-1}\frac{1}{\pi_i} \,
{\ind}_{\{\gamma_0=\lfloor \log_p(y) \rfloor ; \gamma_1=\lceil
  \log_p(\rho(\log_p(y))y) \rceil \}} \right ){\ind}_{\Omega(x,y)}.\nonumber
\end{eqnarray*}
This ends the proof.
\end{proof}

\acknowledgements
\label{sec:ack}
I wish to thank Philippe Robert for his very useful discussions and recommendations, Christine Fricker for her comments on a preliminary version and Mohamed Kamel Eddine Mrad for numerical simulations and graphs.

\bibliographystyle{abbrvnat}
% use the following instead if you encounter problems 
%\bibliographystyle{alpha}
\bibliography{ref}
\label{sec:biblio}

\end{document}